\documentclass[aps,prb,twocolumn,amsmath,amssymb]{revtex4-1}
\usepackage{times}
\usepackage{graphicx}
\usepackage[caption=false]{subfig}
\usepackage{amsfonts}
\usepackage{amsmath, amssymb}
\usepackage{textcomp}
\usepackage{color}
\usepackage{bm}
\usepackage{float}
\usepackage{soul}
\definecolor{DarkRed}{rgb}{0.65,0,0}%
\definecolor{Green}{rgb}{0,0.5,0.3}
\definecolor{Purple}{rgb}{0.3,0,0.65}

\newcommand{\ve}[1]{\boldsymbol{#1}}

\newcommand{\vk}{{\ve{k}}} 
\newcommand{\vech}{\ve{h}} 
\newcommand{\vecf}{\ve{f}} 

\newcommand{\ow}{odd-$\omega$ } 
\newcommand{\ew}{even-$\omega$ } 

\newcommand{\hatx}{\hat{\boldsymbol{x}}} 
\newcommand{\haty}{\hat{\boldsymbol{y}}} 
\newcommand{\hatz}{\hat{\boldsymbol{z}}} 

\newcommand{\e}[1]{\mathrm{e}^{#1}}

\newcommand{\vecsigma}{\boldsymbol{\sigma}}

\newcommand{\veci}{{\ve{i}}}
\newcommand{\vecj}{{\ve{j}}}

\newcommand{\vecn}{\ve{n}}
\newcommand{\vecd}{\ve{d}}
\newcommand{\vecx}{\ve{x}}
\newcommand{\vecy}{\ve{y}}

\newcommand{\vecS}{\ve{S}}

\newcommand{\fd}{f(E_{nk}/2)}

\newcommand{\eg}{\textit{e.g. }}
\newcommand{\etal}{\emph{et al.}}
\def\i{\mathrm{i}}

\begin{document}

\title{Intrinsic Superspin Hall Current}

\author{Jacob Linder}
\thanks{These authors contributed equally to this work}
\author{Morten Amundsen}
\thanks{These authors contributed equally to this work}
\author{Vetle Risingg{\aa}rd}
\affiliation{Department of Physics, NTNU, Norwegian University of Science and Technology, N-7491 Trondheim, Norway}
\affiliation{Center for Quantum Spintronics, Department of Physics, Norwegian University of Science and Technology, NO-7491 Trondheim, Norway}

\begin{abstract}
We discover an intrinsic superspin Hall current: an injected charge supercurrent in a Josephson junction containing heavy normal metals and a ferromagnet generates a transverse spin supercurrent. There is no accompanying dissipation of energy, in contrast to the conventional spin Hall effect. The physical origin of the effect is an antisymmetric spin density induced among transverse modes $k_y$ near the interface of the superconductor arising due to the coexistence of $p$-wave and conventional $s$-wave superconducting correlations with a belonging phase mismatch. Our predictions can be tested in hybrid structures including thin heavy metal layers combined with strong ferromagnets and ordinary $s$-wave superconductors. 
\end{abstract}

\date{\today}

\maketitle

\section{Introduction}

\begin{figure}[b!]
\includegraphics[width=1.0\columnwidth, angle=0]{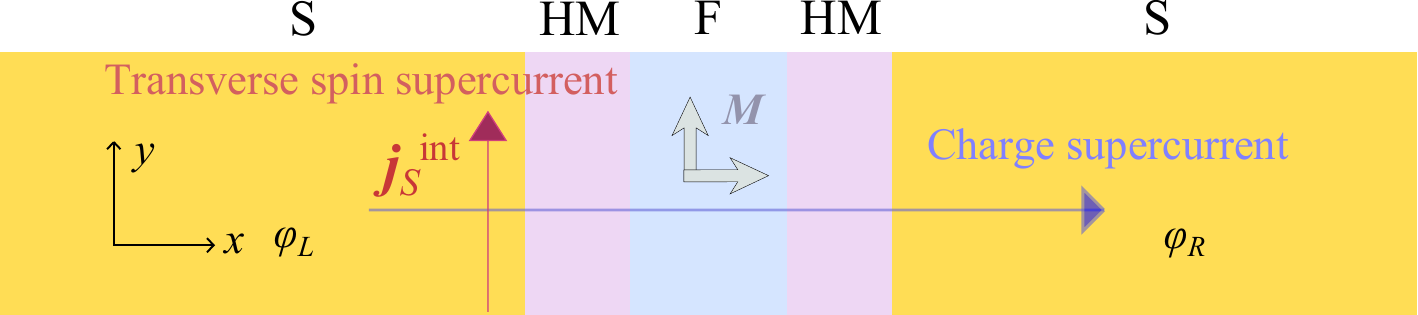}
\caption{(Color online) Suggested experimental setup for demonstration of the superspin Hall current in a Josephson junction. The exchange field in the ferromagnetic region (gray arrows) is directed either along the $\hatx$- or $\haty$-axis. In our calculations, we model the system as a 2D square lattice with periodic boundary conditions in the $\hat{\boldsymbol{y}}$-direction. }
\label{fig:model}
\end{figure}

By combining materials with different properties at the quantum mechanical level into hybrid structures, new physics emerges which often cannot be found in pure materials. The field of superconducting spintronics \cite{linder_nphys_15} is a prime example of this, where the synthesis of superconducting and magnetic correlations have been shown \cite{blamire_jpcm_14, eschrig_rpp_15, beckmann_jpcm_16} to yield physical effects which are interesting both from a fundamental viewpoint and from the viewpoint of potential cryogenic applications. One actively pursued direction in this field has been the prospect of producing dissipationless currents of spin carried by spin-polarized Cooper pairs \cite{keizer_nature_06, eschrig_nphys_08}. The conversion of charge currents to spin currents is known to occur via the spin Hall effect \cite{dyakonov_jetp_71, dyakonov_pla_71, hirsch_prl_99} in conventional spintronics, but is accompanied by dissipation of energy due to the resistive nature of electric currents in non-superconducting structures. Here, we show that it is possible to achieve a dissipationless conversion from charge to spin supercurrents, and vice versa, using conventional superconducting materials. We discover that an injected charge supercurrent in a Josephson junction generates a pure transverse spin supercurrent which thus is time-reversal invariant. Due to the analogy with the conventional spin Hall current, we refer to this as a superspin Hall current. The microscopic origin of the superspin Hall current is a spin magnetization induced at the interface which is antisymmetric in transverse momentum $k_y$. This magnetization is in turn caused by the induction of $p$-wave superconductivity coexisting with conventional spin-singlet pairing. Our predictions can be verified using hybrid structures with thin heavy metal layers combined with strong ferromagnets and ordinary $s$-wave superconductors (see Fig. \ref{fig:model}) and open new vistas for making superconductors compatible with spintronics functionality.

\section{Theory}

In order to describe physics occuring at atomic length scales and also incorporating strong spin-orbit coupling, we use the tight-binding Bogolioubov de Gennes (BdG) framework which is free from the limitations on length-scales and self-energy magnitudes present in \eg quasiclassical theory \cite{chandrasekhar_chapter}. Our Hamiltonian reads:
\begin{align}\label{eq:H}
H &= -t\sum_{\langle \veci,\vecj \rangle \sigma} c_{\veci\sigma}^\dag c_{\vecj\sigma} -\frac{\i}{2}\sum_{\langle \veci,\vecj\rangle\alpha\beta} \lambda_{\veci} c_{\veci\alpha}^\dag \vecn \cdot(\vecsigma \times \vecd_{\veci\vecj})_{\alpha\beta} c_{\vecj\beta}    \notag\\
&- \sum_{\veci\sigma} \mu_{\veci} c_{\veci\sigma}^\dag c_{\veci\sigma}   - \sum_{\veci} U_{\veci} n_{\veci\uparrow}n_{\veci\downarrow} + \sum_{\veci\alpha\beta} c_{\veci\alpha}^\dag (\vech_i\cdot\vecsigma)_{\alpha\beta} c_{\veci\beta}.
\end{align}
Here, $t$ is the hopping integral, $\{c_{\veci\sigma},c^\dag_{\veci\sigma}\}$ are second quantized fermion operators for site $\veci$ and spin $\sigma$, $\vecn$ is a unit vector normal to the interface, $\lambda_{\veci}$ is the site-dependent spin-orbit coupling magnitude, $\vecd_{\veci\vecj} = -\vecd_{\vecj\veci}$ is the nearest-neighbor vector from site $\veci$ to site $\vecj$, $n_{\veci\sigma} = c_{\veci\sigma}^\dag c_{\veci\sigma}$, $\vecsigma$ is the Pauli matrix vector, $\vech_i$ is the local magnetic exchange field, $\mu_\veci$ is the local chemical potential, and $U_\veci$ is the on-site attractive interaction giving rise to superconductivity. For concreteness, we consider a square lattice of size $N_x \times N_y$ with lattice site indices $\veci=(i_x,i_y)$. To demonstrate the superspin Hall current, we consider Fig. \ref{fig:model} which may be experimentally achieved by creating a stack of layers including one magnetic layer (\eg Fe or Co) and two thin heavy metal layers (\eg Pt or Au) sandwiched between two conventional superconductors (\eg Nb or Al). The various terms in Eq. (\ref{eq:H}) exist in their respective regions in Fig. \ref{fig:model}. For instance, the spin-orbit coupling term $\lambda_{\veci}$ is only finite for lattice points inside the heavy metal regions. For brevity of notation, the lattice constant is set to $a=1$ and all length scales are measured relative $a$ whereas all energies are measured relative $t$. Since $\vecn$ is the interface normal ($\vecn=\hat{\boldsymbol{x}}$) the Hamiltonian above is Hermitian without any requirement of symmetrization.  

To simplify the calculations, we assume periodic boundary conditions in the $\hat{\vecy}$-direction, as is common practice \cite{zhu_prb_00, blackschaffer_prb_08, terrade_prb_16}. While this represents an approximation to the geometries considered, it will still allow us to determine the presence of transverse currents. Eq. (\ref{eq:H}) may now be diagonalized by Fourier transforming the fermion operators in the $\hat{\vecy}$-direction; 
\begin{align}
{c_{\veci\sigma} = 1/\sqrt{N_y} \sum_{k_y} c_{i_xk_y\sigma}\e{\i k_y i_y}}. 
\end{align}
Performing a standard mean-field ansatz $\Delta_\veci = -U_\veci \langle c_{\veci\downarrow}c_{\veci\uparrow}\rangle$, one arrives at the Hamiltonian
\begin{align}
H = H_0 + \frac{1}{2} \sum_{ijk} B_{ik}^\dag H_{ijk} B_{jk},
\end{align}
where $H_0$ contains the superconducting condensation energy $N_y\sum_{i \leq N_{x,S}} |\Delta_i|^2/U_i$ 
 (which must be retained when evaluating the free energy of the system). Let $i\equiv i_x$ and $j\equiv j_x$ from now on for brevity of notation. The superconducting region is comprised of $N_{x,S}$ lattice points whereas the heavy metals generating interfacial Rashba spin-orbit coupling and strong ferromagnets have spatial extensions $N_{x,\mathit{HM}}$ and $N_{x,F}$, respectively. The total number of lattice sites is $N_x=N_y$. Setting $k\equiv k_y$, the basis above is 
\begin{align}
B_{ik}^\dag = [c_{ik\uparrow}^\dag\ c_{ik\downarrow}^\dag\ c_{i,-k\uparrow}\ c_{i,-k\downarrow}]^T
\end{align}
and we defined the $4\times4$ matrix:
\begin{align}\label{eq:H2}
H_{ijk} &= \epsilon_{ijk}\hat{\sigma}_0\hat{\tau}_3 +  [h_\veci^y\hat{\sigma}_y+(\lambda\sin k/2)\hat{\sigma}_z]\hat{\tau}_0 +  \notag\\
& + (h_\veci^x\hat{\sigma}_x + h_\veci^z\hat{\sigma}_z)\hat{\tau}_3 + \Delta_i \i\hat{\sigma}_y\hat{\tau}^+ - \Delta_i^* \i\hat{\sigma}_y \hat{\tau}^-.
\end{align}
where 
\begin{align}
\epsilon_{ijk} \equiv -t\cos(k)\delta_{ij} - t(\delta_{i,j+1} + \delta_{i,j-1})/2 - \mu_i\delta_{ij}
\end{align}
and $\hat{\tau}^\pm = \hat{\tau}_1 \pm \i \hat{\tau}_2$. By diagonalizing the above matrix, we end up with the Hamiltonian 
\begin{align}
H = H_0 + \frac{1}{2}\sum_{nk} E_{nk} \gamma_{nk}^\dag \gamma_{nk}, 
\end{align}
where the new (quasiparticle) fermion operators are related to the original ones via the relations 
\begin{align}
c_{ik\uparrow} &= \sum_n u_{ink} \gamma_{nk},\; c_{ik\downarrow} = \sum_n v_{ink} \gamma_{nk},\notag\\
c_{i,-k,\uparrow}^\dag &= \sum_n w_{ink}\gamma_{nk},\; c_{i,-k,\downarrow}^\dag = \sum_n x_{ink}\gamma_{nk}. 
\end{align}
Here, $\{u,v,w,x\}$ are elements of the matrix which diagonalizes the Hamiltonian and are numerically obtained. The diagonalized form of the Hamiltonian makes it trivial to evaluate expectation values of the type $\langle \gamma^\dag_{nk} \gamma_{nk}\rangle = f(E_{nk}/2)$ where $f$ is the Fermi-Dirac distribution function. 

\begin{figure*}[t!]
\includegraphics[width=2.0\columnwidth, angle=0]{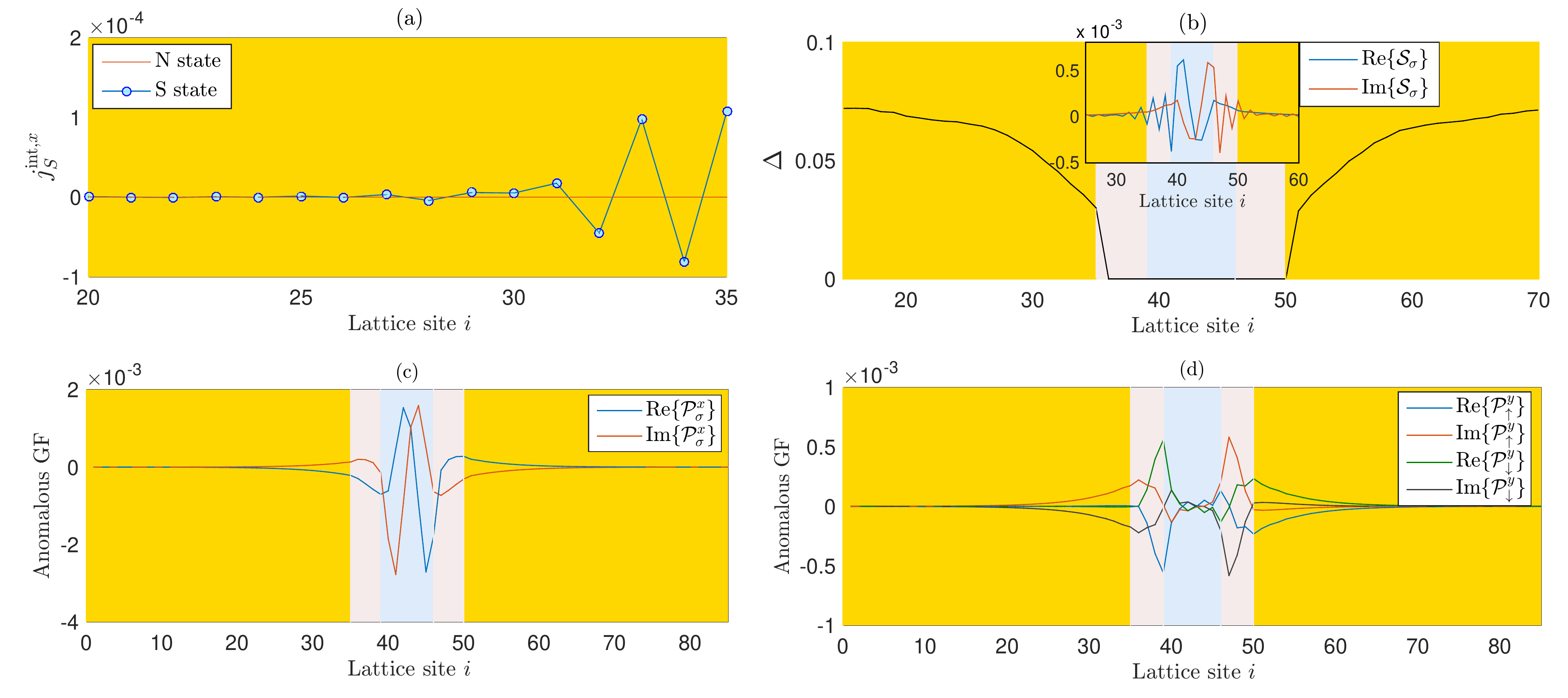}
\caption{(Color online) (a) Superspin Hall current manifested via a transverse spin supercurrent $j^{\text{int},x}_{i,S}$ in the superconducting (S) state. It vanishes in the normal (N) state: $j^{\text{int},x}_{i,S}=0$. (b) Spatial evolution of the superconducting order parameter. \textit{Inset}: $\mathcal{S}_{\sigma,\veci}(\tau)$ with relative time set to $\tau=5$ , (c) $\mathcal{P}^x_{\sigma,\veci}$ and (d) $\mathcal{P}^y_{\sigma,\veci}$. For the inset in panel (b) and in panel (c), the curves are identical for $\sigma=\uparrow$ and $\sigma=\downarrow$. We have used the parameter set specified in the main text, considered the system in Fig. \ref{fig:model}(a), and set $h_y = 0.5$, $\lambda=0.2$, and $\Delta\phi=0.5\pi$.}
\label{fig:main}
\end{figure*}

With the eigenvectors $\{u,v,w,x\}$ and eigenvalues $\{E_{nk}\}$ in hand, we can compute a number of physical quantities in the system under consideration. For instance, the order parameter 
takes the form 
\begin{align}
\Delta_i = -\frac{U_i}{L_y} \sum_{nk}^{\prime} v_{ink} w_{ink}^* [1-\fd]
\end{align}
where the $^{\prime}$ superscript on the sum indicates that only energy eigenvalues $|E_{nk}|<\omega_D$ should be included, and $\omega_D$ is the BCS Debye cut-off frequency. The free energy reads 
\begin{align}
F=H_0 - \frac{1}{\beta}\sum_{nk} \text{ln}(1 + \e{-\beta E_{nk}/2})
\end{align}
whereas the evaluation of charge and spin supercurrent $\vecj_\veci$ and $\vecj_{\veci,S}$ operators requires a consideration of the combined continuity and Heisenberg equation:
\begin{align}
-\nabla \cdot \vecj_\veci = \i[H,\rho_\veci],\; -\nabla\cdot \vecj_\veci^S= \i[H,\vecS_\veci].
\end{align}
Here, 
\begin{align}
\rho_\veci = \sum_\sigma c_{\veci\sigma}^\dag c_{\veci\sigma}
\end{align}
is the charge density operator at site $\veci$ while 
\begin{align}\label{eq:spindensitytotal}
\vecS_\veci = \sum_{\alpha\beta} c_{\veci\alpha}^\dag \vecsigma_{\alpha\beta} c_{\veci\beta}
\end{align}
is the spin density operator (we omitted some constant prefactors such as the electronic charge $|e|$). After a Fourier-transformation, the spin density expectation value at site $i$ reads:
\begin{align}\label{eq:spindensity}
\vecS_i = \sum_{k\alpha\beta} \vecS_{ik},\; \vecS_{ik} = \langle c_{ik\alpha}^\dag \vecsigma_{\alpha\beta} c_{ik\beta} \rangle.
\end{align}
Here, $\boldsymbol{S}_{ik}$ is the momentum-resolved spin density expectation value at lattice point $i$ which will play a prominent role in the discussion later.


A spin supercurrent flowing along the interface has three polarization components and is most conveniently evaluated in the superconducting region: 
\begin{align}\label{eq:spinsuper}
\vecj^\text{int}_{i,S} = \langle \vecj_\veci^S\cdot\hat{\vecy}\rangle = -\frac{8t}{N_y} \sum_{k\alpha\beta} \sin(k)\vecsigma_{\alpha\beta} \langle c_{ik\alpha}^\dag c_{ik\beta}\rangle. 
\end{align}
For instance, the spin supercurrent polarized in the $\hat{\boldsymbol{x}}$- and $\hat{\boldsymbol{y}}$-direction is:
\begin{align}
j^{\text{int},x}_{i,S} &= -\frac{16t}{N_y}\sum_{nk} \sin(k) \text{Re}\{ u_{ink}v^*_{ink}\}f(E_{nk}/2),\notag\\
j^{\text{int},y}_{i,S} &= \frac{16t}{N_y}\sum_{nk} \sin(k) \text{Im}\{ u_{ink}^*v_{ink}\}f(E_{nk}/2).
\end{align}

\section{Results}

\subsection{Superspin Hall current}

We first numerically diagonalize the Hamiltonian given by Eqs. (\ref{eq:H}) and (\ref{eq:H2}) for the Josephson junction shown in Fig. \ref{fig:model} using the parameters $\mu_S=0.9, \mu_N = 0.85, \mu_F = 0.8$, $\omega_D = 0.3$, $N_{x,S} = 35$, $N_{x,HM} = 4$, $N_{x,F} = 7$, $U=2.1$, and $T=0.01$. The order parameter phase is fixed at the last five lattice points in the S regions in order to model supercurrent injection via a phase difference, as is standard in the BdG lattice treatment. Fixing $\Delta\phi=0.5\pi$ gives an effective phase difference between the superconducting interfaces of $\Delta\phi \simeq 0.47\pi$ due to the phase drop inside the superconductors. The following results are not qualitatively sensitive to the parameter choice above. For the above parameter set, and all other sets presented in the figures of this paper, we have checked that the superconducting state minimizes the free energy of the system.

When $\Delta\phi\neq0$, a transverse spin supercurrent appears in the superconducting region as shown in Fig. \ref{fig:main}(a). This demonstrates the intrinsic superspin Hall current. The effect occurs even if one removes one of the heavy metal layers. The spin supercurrent predicted here does not exist in the absence of superconductivity, as also shown in Fig. \ref{fig:main}(a). Reversing the phase difference, $\Delta\phi \to -\Delta\phi$, and thus the charge supercurrent, also reverses the transverse spin supercurrent. Before explaining the microscopic origin of the superspin Hall current, we note that there are both odd- and even-frequency triplet correlations in the system, denoted \ow and \ew from now on. The onsite ($s$-wave) \ow anomalous triplet amplitudes $\mathcal{S}$ are defined as 
\begin{align}
\mathcal{S}_{0,\veci}(\tau) &= \langle c_{\veci\uparrow}(\tau) c_{\veci\downarrow}(0)\rangle + \langle c_{\veci\downarrow}(\tau)c_{\veci\uparrow}(0)\rangle,\notag\\
\mathcal{S}_{\sigma,\veci}(\tau) &= \langle c_{\veci\sigma}(\tau)c_{\veci\sigma}(0)\rangle
\end{align}
where $\tau$ is the relative time coordinate and the subscripts $0$ and $\sigma=\pm1=\uparrow,\downarrow$ denote the spin projection along the quantization axis. All $\mathcal{S}$ vanish at $\tau=0$. The $p$-wave \ew anomalous triplet amplitudes $\mathcal{P}$ have both a $p_x$ and $p_y$ wave component. They are defined as 
\begin{align}
\mathcal{P}^{x(y)}_{0,\veci} &= \sum_\pm \pm(\langle c_{\veci\uparrow} c_{\veci\pm\hat{\vecx}(\hat{\vecy}),\downarrow}\rangle + \langle c_{\veci\downarrow} c_{\veci\pm\hat{\vecx}(\hat{\vecy}),\uparrow}\rangle),\notag\\
\mathcal{P}^{x(y)}_{\sigma,\veci} &= \sum_\pm \pm \langle c_{\veci\sigma}c_{\veci\pm\hat{\vecx}(\hat{\vecy}),\sigma} \rangle
\end{align}
The existence of these correlations and their spatial distribution throughout the system is shown in Fig. \ref{fig:main}(b)--\ref{fig:main}(d), proving how they arise precisely near the interfaces between the superconductor and heavy metals where the transverse spin supercurrent flows. The triplet components of the Cooper pairs are generated from the broken spin rotational symmetry in our system whereas the $p$-wave orbital symmetry emerges as a result of broken translational symmetry due the presence of interfaces \cite{tanaka_prl_07, eschrig_jltp_07} and due to the presence of spin-orbit interactions. Note how the quantities pairing amplitudes $\mathcal{S}$ and $\mathcal{P}$ are by definition $\vk$-independent. The $\vk$-resolved anomalous Green functions, which are odd under $\vk\to(-\vk)$ for \eg $p$-wave pairing, will be examined in the following subsection as they play an important role in understanding the appearance of a transverse spin supercurrent.

\begin{figure*}[t!]
\includegraphics[width=2.0\columnwidth, angle=0]{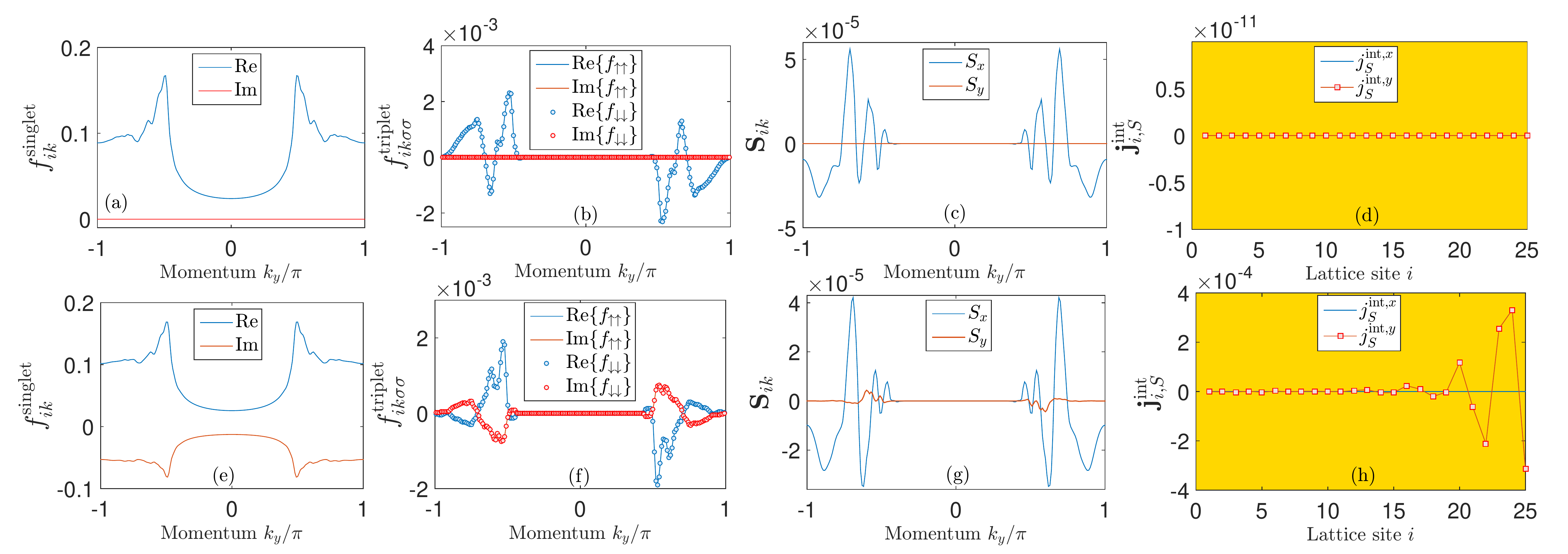}
\caption{(Color online) \textit{Upper row}: The panels show the induced singlet anomalous Green function $f^\text{singlet}_{ik}$, the triplet correlations $f_{ik\sigma\sigma}^\text{triplet}$ evaluated at lattice point $i_x=N_S-1$ (right at the S/N interface), the induced spin-magnetization density $\boldsymbol{S}_{ik}$ evaluated at lattice point $i_x=N_S-1$ (right at the S/N interface), and the superspin Hall current $\boldsymbol{j}^\text{int}_{i,S}$ from left to right. Here, the phase difference has been set to $\Delta\phi=0$ while $h=0.3, \lambda=0.3$, $T=0.005$, $N_S=25, N_N=4, N_F=5$, $N_y=200$, and $\boldsymbol{h}=h\hat{\boldsymbol{x}}$. \textit{Lower row}: Same as the upper row except that $\Delta\phi=\pi/2$. In this case, the coexistence of triplet and singlet correlations which are phase mismatched produce an antisymmetric spin density $\boldsymbol{S}_{ik}$ which in turn gives rise to a finite superspin Hall current. In the line labels of the second panels from the left, we have abbreviated $f_{ik\sigma\sigma}^\text{triplet} \equiv f_{\sigma\sigma}$. In this figure, the exchange field has been rotated to the $\hatx$-direction to show that the superspin Hall current exists also in this case. In all other plots in this manuscript, the exchange field points in the $\haty$-direction.}
\label{fig:anomalous}
\end{figure*}

The transverse spin-supercurrent in the present system exists when the exchange field contribution $\vech \cdot \vecsigma$ to the Hamiltonian does not commute with the spin-orbit contribution $\lambda \sin(k)\sigma_z$. In effect, the superspin Hall current arises when:
\begin{align}
[\vech\cdot\vecsigma, \lambda \sin(k)\sigma_z] \neq 0.
\end{align}
This means that the exchange field must be oriented in the $xy$-plane of the system shown in Fig. \ref{fig:model}. If the exchange field is oriented along the $z$-axis, no superspin Hall current exists.

The polarization of the transverse spin-supercurrent is also dictated by the orientation of the exchange field $\vech$. Comparison of Figs. \ref{fig:main}(a) and \ref{fig:anomalous}(h) shows that the spin-supercurrent polarization is perpendicular to $\vech$.

\subsection{Microscopic origin}

To explain the physical origin of the superspin Hall current in the system, we first note the close relation between the spin magnetization and the spin supercurrent in the system. From Eqs. (\ref{eq:spindensity}) and (\ref{eq:spinsuper}), the only difference between them is a factor $\sin(k_y)$ inside the summation. If the momentum-resolved spin magnetization $\vecS_{ik}$ is antisymmetric in momentum $k_y$, it will vanish when summed over the momentum index. However, due to the extra factor $\sin(k_y)$ an antisymmetric spin magnetization gives a symmetric spin supercurrent, which is thus finite upon summation over $k_y$. The factor $\sin(k_y)$ and the resulting difference in symmetry are physically reasonable. If a spin density is antisymmetric in momentum $k_y$, there will exist a net spin flow since the spin current requires an extra multiplication with the group velocity 
\begin{align}
\partial \epsilon_\vk/\partial k_y \propto \sin(k_ya)
\end{align}
for each transverse mode. On the other hand, a spin density that is symmetric in $k_y$ does not induce any spin current. 

An antisymmetric spin density in the momentum index $k_y$ may emerge whenever conventional superconducting singlet pairing and triplet pairing (such as $p_y$-wave) coexists, for instance near interfaces, as we will explain below. 
A general superconducting order parameter $F_{i\vk}$ accounting for both singlet and triplet pairing (considering here the \ew symmetry contribution) can be written as
\begin{align}
F_{i\vk} = (f_{i,s} + \vecf_{i,\vk}\cdot\vecsigma)\i\sigma_2
\end{align}
where $f_{i,s}$ is the singlet component and $\vecf_{i,\vk} = -\vecf_{i,-\vk}$ is a vector containing the triplet components according to 
\begin{align}\label{eq:dvector}
\vecf = \frac{1}{2}[f_{\downarrow\downarrow}-f_{\uparrow\uparrow}, -\i(f_{\downarrow\downarrow}+f_{\uparrow\uparrow}), 2f_{\uparrow\downarrow}].
\end{align}
Above, we suppressed the $(i,\vk)$-indices on the triplet anomalous Green functions $f_{i\vk\sigma\sigma'}$ for brevity of notation and we also do so below when the index is not of importance for the argument. A non-unitary superconducting state, where the Cooper pairs have a finite spin expectation value, is defined by $FF^\dag$ not being proportional to the unit matrix. A straightforward calculation shows that
\begin{align}
FF^\dag = |f_s|^2 + |\vecf_\vk|^2 + \vecsigma\cdot[(f_s\vecf_{\vk}^* + f_s^*\vecf_{\vk}) + \i(\vecf_\vk \times \vecf_\vk^*)].
\end{align}
The term $\i(\vecf_\vk \times \vecf_\vk^*)$ determines the spin expectation value of pure triplet Cooper pairs whereas $(f_s\vecf_{\vk}^* + f_s^*\vecf_{\vk})$ determines the spin-magnetization of a given mode $k_y$ resulting from the coexistence of singlet and triplet pairing. The spin magnetization arising due to the Cooper pairs in the system thus in general takes the following form for a given mode $k_y$:
\begin{align}
\boldsymbol{S}^\mathrm{Cooper}_\vk \propto (f_s\vecf_{\vk}^* + f_s^*\vecf_{\vk}) + \i(\vecf_\vk \times \vecf_\vk^*)
\end{align}
Performing a summation over modes $k_y$, one obtains the total spin density. Therefore, it is clear that if $(f_s\vecf_{\vk}^* + f_s^*\vecf_{\vk}) = 2\text{Re}\{f_s\vecf_\vk^*\}$ is non-zero, it will be antisymmetric in $k_y$ due to the fundamental property of the triplet vector $\vecf_\vk$. It is crucial to note that the existence of $p$-wave \textit{triplet pairing alone is not sufficient} to produce an antisymmetric spin density in $k_y$-space. First of all, it has to coexist with singlet pairing. But even such a scenario is not sufficient, as it is only the real part of the product $f_s\vecf_\vk^*$ that contributes. Consider for instance the case where singlet pairing coexists with $S_z=0$ triplet pairing, such that $\vecf_\vk \parallel \hatz$. According to our above argumentation, this should produce a magnetization in the $\hatz$ direction. It is not immediately obvious how a magnetization in the $\hatz$ direction can arise from singlet pairs (which are spinless) and triplet pairs with zero spin projection along the $\hatz$ axis. Therefore, we provide a detailed exposition of the physical mechanism behind this effect in the Appendix.

With this in mind, we can now explain why the superspin Hall current appears. As argued above, this current will exist when an antisymmetric spin density is induced near the interface. The spin density, in turn, is determined by the generation of $p$-wave superconducting correlations coexisting with conventional singlet ones when these have an appropriate relative phase such that $\text{Re}\{f_s\vecf_\vk^*\}\neq0$ (as explained in the Appendix). The equal spin-pairing triplet anomalous Green functions may be obtained as
\begin{align}
f_{ik\sigma\sigma}^\text{triplet} &= \langle c_{i,k,\sigma} c_{i,-k,\sigma}\rangle \notag\\
&= \begin{cases}
\sum_n u_{ink} w_{ink}^*[1-f(E_{nk}/2)] \text{ for $\sigma=\uparrow$}\\
\sum_n v_{ink} x_{ink}^*[1-f(E_{nk}/2)] \text{ for $\sigma=\downarrow$},
\end{cases}
\end{align}
where, as before, we have used the shorthand notation of $k_y\equiv k$.
We now illustrate two instructive cases in Fig. \ref{fig:anomalous}. In the upper row (a--d), the phase-difference is $\Delta\phi=0$ (no current injected) while in the lower row (e--h) the phase-difference is $\Delta\phi=\pi/2$ (finite current injected). In both cases, we have set $h=0.3, \lambda=0.3$ and $\vech = h\hat{\boldsymbol{x}}$. We also chose a different system size, exchange field orientation, and number of transverse modes than in the previous figures in order to show that the effect does not depend on these details: $N_S=25, N_N = 4, N_F = 5$ and $N_y=200$. As expected, a finite net magnetization $S_x$ exists in the upper row which comes from the inverse proximity effect caused by the magnetic region. However, there exists no net or $k_y$-resolved magnetization $S_y$ despite the fact that the anomalous triplet correlations $f_{ik\sigma\sigma}^\text{triplet} \equiv f_{\sigma\sigma}$ are non-zero. The reason for this is that they are purely imaginary, as seen in the figure. Consequently, $\text{Re}\{f_s\vecf_\vk^*\}=0$ since the singlet ones are purely real in the absence of a phase-gradient. Note how the figure shows that $f_{\uparrow\uparrow} = f_{\downarrow\downarrow}$, such that no antisymmetric contribution is made to the $\boldsymbol{x}$-component according to Eq. (\ref{eq:dvector}). The finite magnetization induced along the $\boldsymbol{x}$-direction is instead caused by the \ow triplet component. In general, the triplet vector $\vecf$ can have both a symmetric term in $\vk$ (the \ow component) and an antisymmetric term in $\vk$ (the \ew component). Only the latter contributes to the spin supercurrent in the present context, as explained above.

\begin{figure}[t!]
\includegraphics[width=1.0\columnwidth, angle=0]{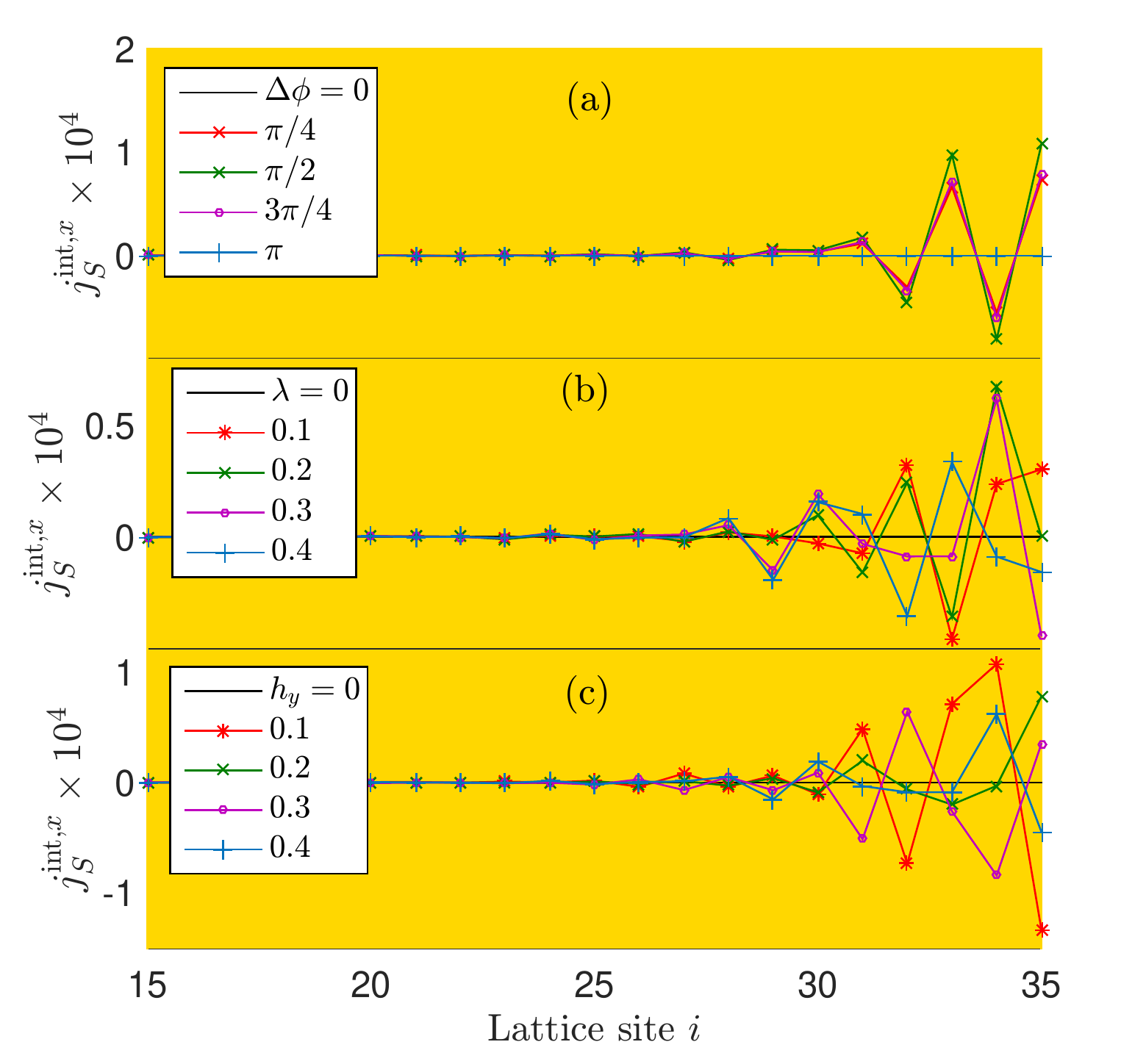}
\caption{(Color online) The dependence of the superspin Hall current $j^{\text{int},x}_{i,S}$ on physical parameters in the system [Fig. \ref{fig:model}(a)]. (a) $\lambda=0.2$ and $h_y=0.5$ for various values of the phase difference $\Delta\phi$. (b) $\Delta\phi=\pi/2$ and $h_y=0.4$ for several spin-orbit magnitudes $\lambda$. (c) $\lambda=0.3$ and $\Delta\phi=\pi/2$ for various values of the exchange field $h_y$. The value of the remaining parameters are the same as in Fig. \ref{fig:main}. The background color indicates in which region the current has been evaluated [compare with left part of Fig. \ref{fig:model}(a)].}
\label{fig:dependence}
\end{figure}

\begin{figure}[b!]
\includegraphics[width=1.0\columnwidth, angle=0]{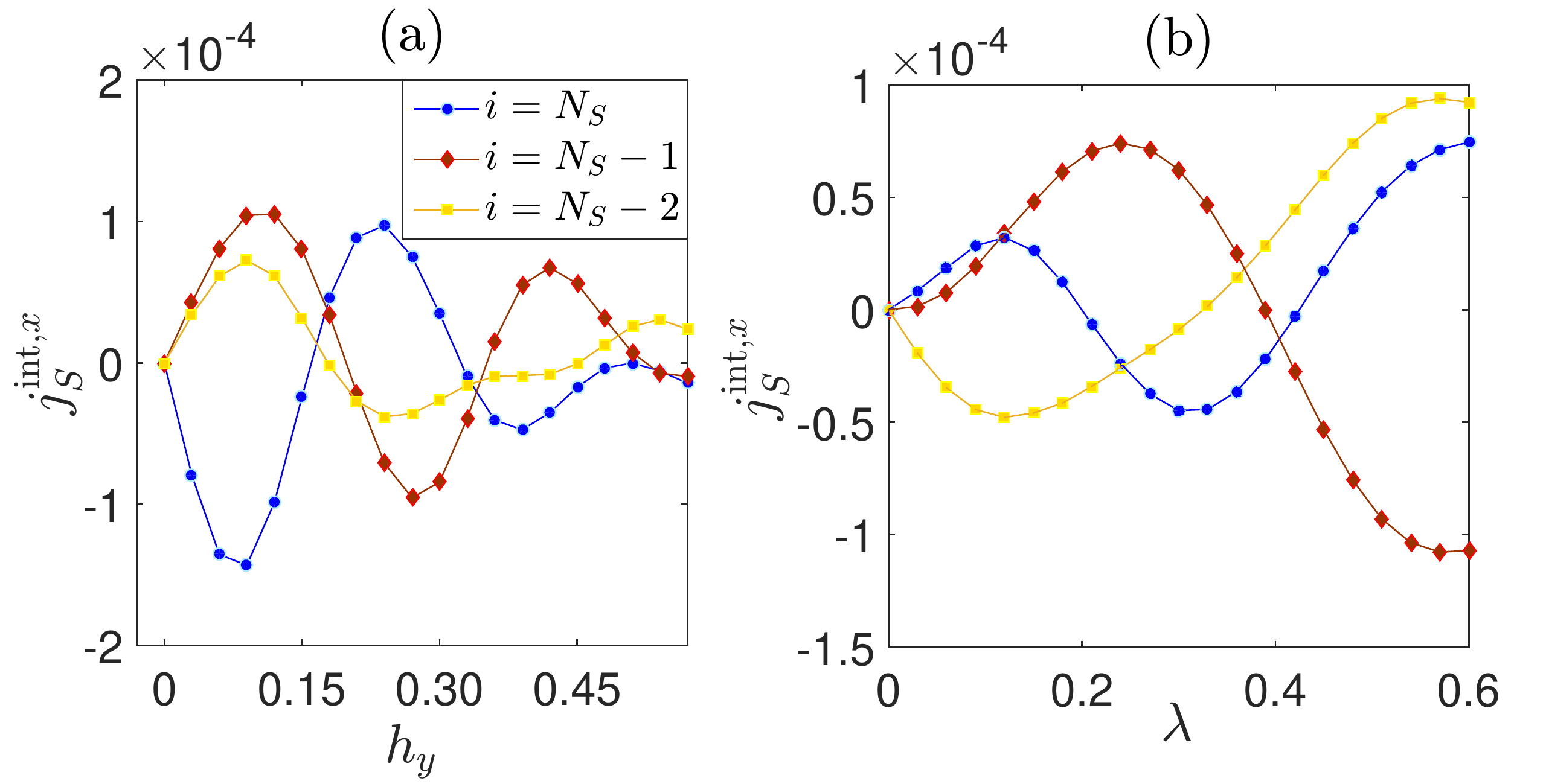}
\caption{(Color online) The dependence of the superspin Hall current $j^{\text{int},x}_{i,S}$ on physical parameters in the system, evaluated at different lattice sites $i$. We have set $\Delta\phi=0.5\pi$ and (a) $\lambda=0.3$ and (b) $h_y=0.4$. The value of the remaining parameters are the same as in Fig. \ref{fig:main}. An scattering potential $V_\text{int}=0.1$ at each of the interfaces was also added here to show that the effect is resilient toward interfacial scattering. The current oscillates with both $h$ and $\lambda$ and eventually decays with both as these quantities increase and suppress superconductivity.}
\label{fig:josephsonlambdah}
\end{figure}

Consider now instead the lower row where a finite phase-difference exists. The singlet and triplet correlations are now complex because of $\Delta\phi\neq0$, and as a result the $\boldsymbol{y}$-component of the spin-magnetization [which exists since the term $\text{Re}\{f_s\vecf_\vk^*\}$ is non-zero] is finite and antisymmetric in $k_y$. Although no net magnetization exists in the $\boldsymbol{y}$-direction, \textit{a net spin supercurrent now exists} due to the relation between Eqs. (\ref{eq:spindensity}) and (\ref{eq:spinsuper}) explained above. A phase-gradient is thus physically required in order to render the singlet and triplet $p_y$-wave correlations complex: otherwise, no antisymmetric spin magnetization associated with a non-unitary superconducting state exists and the spin supercurrent is zero. This explains the origin of the superspin Hall current predicted in this manuscript. 

The above explanation is consistent irrespective of the direction of the in-plane exchange field. For instance, if we instead choose $\vech = h\hat{\boldsymbol{y}}$ one finds that the triplet $p_y$-wave anomalous function is purely real at $\Delta\phi=0$ and that $f_{\uparrow\uparrow} =-f_{\downarrow\downarrow}$. In this case, there is no contribution to the $\hat{\boldsymbol{y}}$-component according to Eq. (\ref{eq:dvector}) and although $\vecf_\vk\cdot\hat{\vecx} \neq 0$ there is still no antisymmetric spin density since $\text{Re}\{f_s\vecf_\vk^*\}=0$. If $\Delta\phi\neq 0$, on the other hand, $\text{Re}\{f_s\vecf_\vk^*\}$ is finite in the $\hat{\boldsymbol{x}}$-direction and a spin supercurrent polarized in this direction appears as seen in the figure. 

The spatial dependence of the superspin Hall current on the phase difference, the Rashba spin-orbit interaction, and exchange field is shown in Fig \ref{fig:dependence}. The effect vanishes both in the absence of superconductivity $(\Delta\phi=0)$ and in the absence of a charge supercurrent $(\Delta\phi=0)$, as follows from the above explanation of the physical origin of the effect. We also find that the magnitude of the transverse current $j^{\text{int},x}_{i,S}$ evaluated at the superconducting interface ($i=N_S\equiv N_{x,S}$) oscillates with both the Rashba strength and the magnitude of the exchange field for the parameter regimes we have investigated, as shown in Fig. \ref{fig:josephsonlambdah}. The effect is also purely sinusoidal as a function of the superconducting phase difference $\Delta\phi$ (not shown). The oscillations could stem from the change in $\vk$-space band structure to the inverse proximity effect near the interface as one varies the magnitude of $h$ and $\lambda$, as the detailed $\vk$-dependence of the spin magnetization (and thus in turn magnitude of the spin supercurrent after summation over $\vk$) will be affected by the details of the band structure.

\begin{figure}[H]
\includegraphics[width=1.0\columnwidth, angle=0]{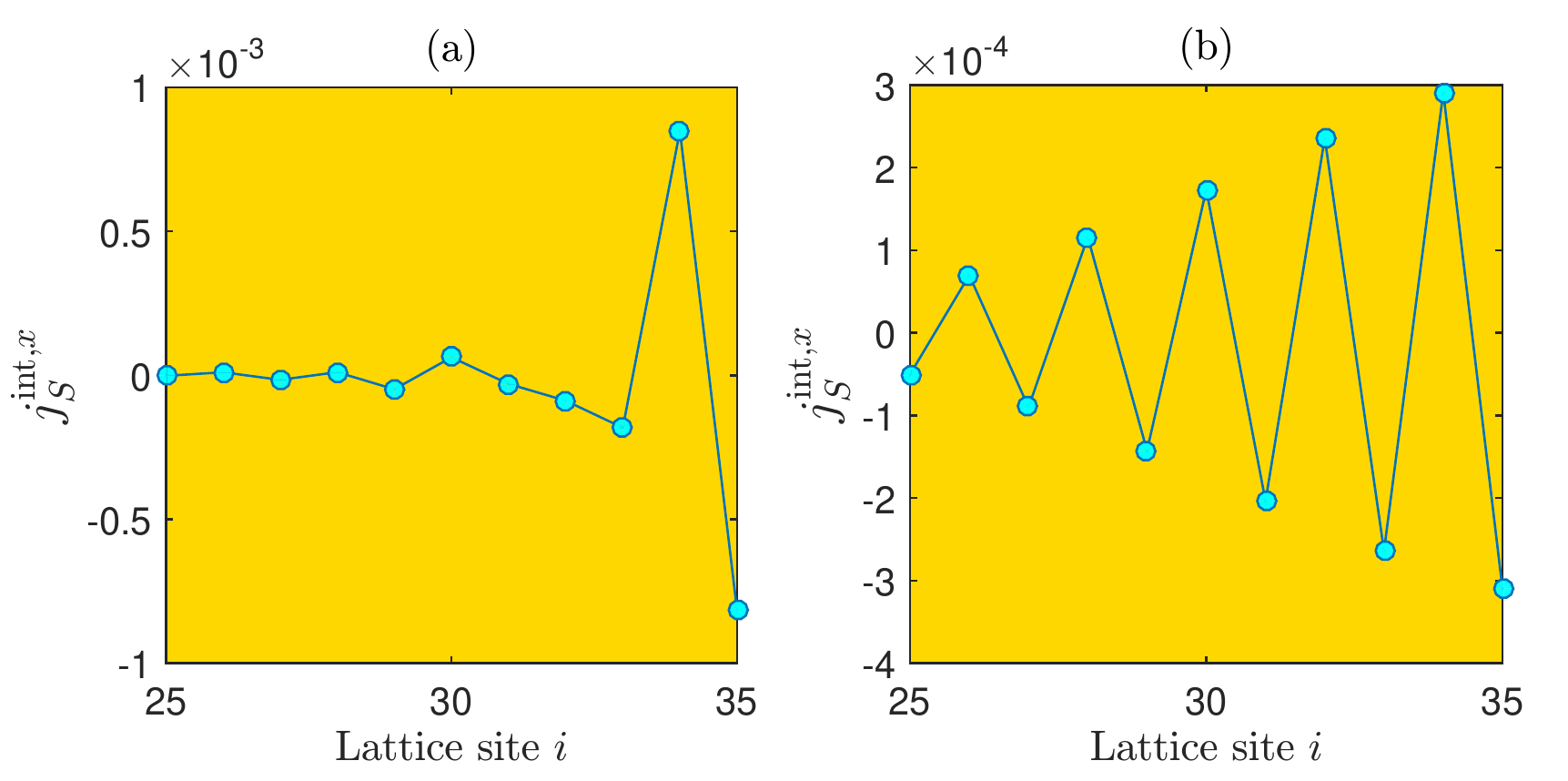}
\caption{(Color online) Change in oscillation length of the superspin Hall current as the magnitude of the Rashba spin-orbit parameter is altered. The plots show the cases (a) $\lambda=0.1$ and (b) $\lambda=2.1$. The exchange field is $\boldsymbol{h} = h\hat{\boldsymbol{y}}$ with $h=0.3$, $T=0.005$, and the other parameters are the same as in Fig. \ref{fig:main}.}
\label{fig:soc}
\end{figure}

The atomic-scale superimposed oscillations are characteristic for physical quantities in ballistic quantum mechanical systems and are also present in \eg the proximity-induced magnetization in conventional superconductors \cite{halterman_prb_08} and helical edge-mode currents in triplet superconductors \cite{terrade_prb_16}. It should be noted, however, that the oscillation period of the spin supercurrent here depends on the system parameters. This is shown in Fig. \ref{fig:soc} where it is clear that the oscillation period is altered by changing the magnitude of the Rashba parameter. The origin of the oscillations is likely to similar to that described in Ref. \onlinecite{terrade_prb_16}, namely due to an interplay between the renormalized spectral weight in the superconductor due to the inverse proximity effect and how the $p$-wave superconducting correlations decay as a result.

\section{Concluding remarks}

Previous theoretical work has considered spin accumulation from spin Hall effects in superconducting structures \cite{pandey_prb_12, kontani_prl_09, sengupta_prb_06, bergeret_prb_16, malshukov_prb_10, malshukov_prb_17} and a recent experimental work \cite{wakamura_natmat_15} demonstrated an enhancement of the inverse spin Hall signal \cite{hirsch_prl_99} in a superconductor by three orders of magnitude. A similar edge spin magnetization might occur from the superspin Hall current predicted in this work. Although the interface between a superconductor and a ferromagnet breaks inversion symmetry on its own, the purpose of the HM layers is to enhance the magnitude of the resulting Rashba interaction. A transverse spin current induced by a charge supercurrent was also considered in Ref. \onlinecite{zhihong_cpb_12}, albeit in a different setup where spin-orbit coupling was present in the entirety of one superconducting region and where no magnetism was present.  Ref. \onlinecite{malshukovchu} considered spin Hall effects in a Josephson setup both with and without an electric bias voltage applied to the system. 


It is worth remarking that in comparison to the typical spin Hall phenomenology, where an injected current in the $x$-direction is deflected in the $y$-direction and polarized in the $z$-direction, the spin supercurrent here is not polarized perpendicularly to the plane defined by its injection and deflection direction. However, similarly to the conventional spin Hall phenomenology, the spin supercurrent arises as a direct consequence of Cooper pairs that are polarized in the $z$-direction. The details regarding how $S_z=\pm 1$ Cooper pairs give rise to a spin supercurrent polarized in the $xy$-plane have been covered in detail in the main body of this manuscript.

Interesting future directions to explore include the precise circulation pattern of the superspin Hall current predicted herein in a finite-width sample and the possible accompanying edge spin accumulation due to triplet Cooper pairs.

\acknowledgments

J.L. thanks J. A. Ouassou for helpful discussions regarding the numerical analysis. The authors thank J. A. Ouassou, S. Jacobsen, A. Black-Schaffer, and H. Simensen for discussions. Funding via the “Outstanding Academic Fellows” program
at NTNU, the NV-Faculty, and
the Research Council of Norway Grant numbers 216700 and
240806 and funding for the Center of Excellence QuSpin, is gratefully acknowledged.

\appendix

\section{Magnetization arising out of a non-unitary coexistence of singlet and triplet pairing}\label{sec:app}

Consider for simplicity a bulk system where singlet pairing $\Delta_s$ coexists with $S_z=0$ triplet pairing $\Delta_{\vk} = -\Delta_{-\vk}$. The Hamiltonian reads
\begin{align}
H = \sum_\vk \phi_\vk^\dag M_\vk\phi_\vk
\end{align}
where we used a basis $\phi_\vk = (c_{\vk\uparrow}, c_{\vk\downarrow}, c_{-\vk\uparrow}^\dag, c_{-\vk\downarrow}^\dag)^\text{T}$ and defined
\begin{align}
M_\vk = \begin{pmatrix}
\epsilon_\vk & 0 & 0 & \Delta_\vk + \Delta_s \\
0 & \epsilon_\vk & \Delta_\vk - \Delta_s & 0 \\
0 & \Delta_\vk^* - \Delta_s^* & -\epsilon_\vk & 0 \\
\Delta_\vk^* + \Delta_s^* & 0 & 0 & -\epsilon_\vk \\
\end{pmatrix}
\end{align}
The four eigenvalues are given as $\left\{E_+,E_-,-E_+,-E_-\right\}$, where
\begin{align}
E_\pm = \sqrt{\epsilon_\vk^2 + |\Delta_s \pm \Delta_\vk|^2}.
\end{align}
Performing a standard diagonalization of the Hamiltonian by introducing a new quasiparticle basis
\begin{align}
\gamma_\vk = (\gamma_{1\vk},\gamma_{2\vk},\gamma_{3\vk},\gamma_{4\vk})^\text{T},
\end{align}
where $\gamma_{i\vk}$ are second-quantized fermion operators, one arrives at
\begin{align}
H &= \sum_\vk [E_{\vk+} (\gamma_{1\vk}^\dag \gamma_{1\vk} - \gamma_{2\vk}^\dag\gamma_{2\vk}) \notag\\
&+ E_{\vk-} (\gamma_{3\vk}^\dag \gamma_{3\vk} - \gamma_{4\vk}^\dag\gamma_{4\vk})].
\end{align}
The relation between the original fermion operators $c$ and the new ones $\gamma$ is
\begin{align}\label{eq:transform}
\phi_\vk = P_\vk\gamma_\vk
\end{align}
where $P_\vk$ is the diagonalizing matrix containing the eigenvectors of the original Hamiltonian
\begin{align}
P_\vk = \begin{pmatrix}
g_+(E_{\vk+}) & g_+(-E_{\vk+}) & 0 & 0 \\
0 & 0 & g_-(E_{\vk-}) & g_-(-E_{\vk-}) \\
0 & 0 & 1 & 1 \\
1 & 1 & 0 & 0 \\
\end{pmatrix}
\end{align}
and we defined the auxiliary quantity
\begin{align}\label{eq:g}
g_\pm(E) = \frac{\Delta_\vk\pm\Delta_s}{E-\epsilon_\vk}.
\end{align}
Now, the magnetization of the system in the $\hatz$-direction is computed according to Eq. (\ref{eq:spindensity}):
\begin{align}\label{eq:spinapp}
S_z = \sum_{\vk\sigma} \sigma \langle c_{\vk\sigma}^\dag c_{\vk\sigma}\rangle.
\end{align}
To see how this magnetization is directly influenced by the coexistence of singlet and triplet pairs in the system, we replace the original fermion operators in Eq. (\ref{eq:spinapp}) with the new ones according to Eq. (\ref{eq:transform}). Considering for simplicity the $T=0$ limit, one arrives at
\begin{align}\label{eq:spinexpression}
S_z = \sum_\vk [|g_+(-E_{\vk+})|^2 - |g_-(-E_{\vk-})|^2].
\end{align}
At this point, we distinguish between unitary and non-unitary states. In the \textit{unitary case}, we have Re$\{\Delta_s\Delta_\vk^*\}=0$ so that $E_{\vk+} = E_{\vk-}$: the magnitude of both gaps $\Delta_\pm = \Delta_\vk \pm \Delta_s$ is equal. Moreover, it follows from Eq. (\ref{eq:g}) that in the unitary case one has $|g_+(x)| = |g_-(x)|$. Combining these two facts, it follows that the term inside the summation $\sum_\vk$ in Eq. (\ref{eq:spinexpression}) is zero for any $\vk$-value. In effect, there is no magnetization at any $\vk$-point and obviously no net magnetization either. \\

Consider now instead a \text{non-unitary state} where Re$\{\Delta_s\Delta_\vk^*\} \neq 0$. In this case, the magnitudes of the gaps $\Delta_\pm$ are \textit{different}. Now, the term inside the summation of Eq. (\ref{eq:spinexpression}) is no longer zero for a given $\vk$-point. In effect, there exists a $\vk$-resolved magnetization. The \textit{total magnetization}, obtained after a summation over $\vk$, is nevertheless zero even in the non-unitary case. This can be verified by splitting the sum in Eq. (\ref{eq:spinexpression}) into $\vk>0$ and $\vk<0$ (the contribution from $\vk=0$ vanishes) and using the general relation $E_{\vk,+} = E_{-\vk,-}$. 

The above derivation establishes mathematically why a $\vk$-resolved, antisymmetric spin magnetization exists when singlet and $p$-wave triplet pairing coexists in a non-unitary state, precisely as in the system considered in the main body of this manuscript. The \textit{physical picture} can be understood by going back to the fact that there exists two gaps with different magnitude in the system. It is well-known that the superconducting order parameter (gap) determines the condensation energy and binding energy between the electrons comprising the Cooper pairs. In particular, the Cooper pair density is proportional to the square of the magnitude of the gap. The point here is that Cooper pairing between two electron states $|\vk,\uparrow\rangle$ and $|-\vk,\downarrow\rangle$ are associated with a gap magnitude $|\Delta_\vk + \Delta_s|\equiv |\Delta_+|$ whereas pairing between two electron states $|\vk,\downarrow\rangle$ and $|-\vk,\uparrow\rangle$ are associated with a different gap magnitude $|\Delta_\vk - \Delta_s| \equiv |\Delta_-|$. This can be seen directly from the Hamiltonian which contains the terms $c_{\vk\uparrow}^\dag c_{-\vk\downarrow}^\dag \Delta_+$ and $c_{\vk\downarrow}^\dag c_{-\vk\uparrow}^\dag \Delta_-$. Now, if $|\Delta_+|>|\Delta_-|$ for a given $\vk$-value it is clear that the system will favor Cooper pairs where the $\uparrow$-electron of the pair sits at $\vk$ whereas the $\downarrow$-electron sits at $-\vk$, since the Cooper pair state where the $\uparrow$-electron sits at $-\vk$ and the $\downarrow$-electron sits at $\vk$ has a smaller binding energy. Therefore, a \textit{net spin magnetization arises at $\vk$} since there exists a surplus of $\uparrow$-spins there compared to $\downarrow$-spins due to the difference in Cooper pair density stemming from the different gap magnitudes. Simultaneously, the opposite magnetization arises at $-\vk$ since at that momentum the situation is reversed: $|\Delta_-|$ is larger than $|\Delta_+|$ at $-\vk$. 

In this way, the different magnitudes of the two gaps in a system where singlet pairing coexists with $S_z=0$ triplet pairing in a non-unitary state cause the Cooper pairs to provide a $\vk$-resolved magnetization in the $\hatz$-direction despite the fact that the net Cooper pair spin in the $\hatz$-direction is zero.


\begin{thebibliography}{999}


\bibitem{linder_nphys_15}  J. Linder and J. W. A. Robinson. Superconducting spintronics. Nat. Phys. \textbf{11}, 307 (2015).

\bibitem{ryazanov_prl_01} V. V. Ryazanov, V. A. Oboznov, A. Yu. Rusanov, A. V. Veretennikov, A. A. Golubov, and J. Aarts. Coupling of Two Superconductors through a Ferromagnet: Evidence for a $\pi$-junction. Phys. Rev. Lett. \textbf{86}, 2427 (2001).

\bibitem{blamire_jpcm_14} M. G. Blamire and J. W. A. Robinson. The interface between superconductivity and magnetism: understanding and device prospects. J. Phys.: Condens. Matter \textbf{26}, 453201 (2014).

\bibitem{eschrig_rpp_15} M. Eschrig. Spin-polarized supercurrents for spintronics: a review of current progress. Rep. Prog. Phys. \textbf{78}, 104501 (2015).

\bibitem{beckmann_jpcm_16} D. Beckmann. Spin manipulation in nanoscale superconductors. J. Phys.: Condens. Matter \textbf{28}, 163001 (2016).

\bibitem{keizer_nature_06} R. S. Keizer \etal. A spin triplet supercurrent through the half-metallic
ferromagnet CrO$_2$. Nature \textbf{439}, 825 (2006).

\bibitem{eschrig_nphys_08} M. Eschrig and T. L{\"o}fwander. Triplet supercurrents in clean and disordered
half-metallic ferromagnets. Nature Phys. \textbf{4}, 138 (2008)


\bibitem{dyakonov_jetp_71} M. I. Dyakonov and V. I. Perel. Possibility of orientating electron spins with current. Sov. Phys. JETP Lett. \textbf{13}, 467 (1971).

\bibitem{dyakonov_pla_71} M.I. Dyakonov and V.I. Perel. Current-induced spin orientation of electrons in semiconductors. Phys. Lett. A. \textbf{35}, 459 (1971).

\bibitem{hirsch_prl_99} J. E. Hirsch. Spin Hall Effect. Phys. Rev. Lett. \textbf{83}, 1834 (1999).

\bibitem{chandrasekhar_chapter} V. Chandrasekhar. \textit{Proximity-Coupled Systems: Quasiclassical Theory of Superconductivity in Superconductivity:
Conventional and Unconventional Superconductors} (eds. Bennemann, K. H. \& Ketterson, J. B.) Ch. 8, 279-313 (Springer Berlin Heidelberg, 2008).

\bibitem{zhu_prb_00} J.-X. Zhu and C. S. Ting. Proximity effect, quasiparticle transport, and local magnetic moment in ferromagnet–
$d$-wave-superconductor junctions. Phys. Rev. B \textbf{61}, 1456 (2000).

\bibitem{blackschaffer_prb_08} A. M. Black-Schaffer and S. Doniach. Self-consistent solution for proximity effect and Josephson current in ballistic graphene SNS Josephson junctions. Phys. Rev. B \textbf{78}, 024504 (2008).

\bibitem{terrade_prb_16} D. Terrade, D. Manske, and M. Cuoco. Control of edge currents at a ferromagnet–-triplet superconductor interface by multiple helical modes. Phys. Rev. B. \textbf{93}, 104523 (2016). 

\bibitem{halterman_prb_08} K. Halterman, O. T. Valls, and P. H. Barsic. Induced triplet pairing in clean $s$-wave superconductor/ferromagnet layered structures. Phys. Rev. B \textbf{77}, 174511 (2008).

\bibitem{tanaka_prl_07} Y. Tanaka, A. A. Golubov, S. Kashiwaya, and M. Ueda. Anomalous Josephson Effect between Even- and Odd-Frequency Superconductors. Phys. Rev. Lett. \textbf{99}, 037005 (2007).

\bibitem{eschrig_jltp_07} M. Eschrig, T. L{\"o}fwander, T. Champel, J. C. Cuevas, J. Kopu, Gerd Sch{\"o}n. Symmetries of Pairing Correlations in Superconductor-Ferromagnet Nanostructures. J. Low Temp. Phys. \textbf{147}, 457 (2007).

\bibitem{pandey_prb_12} S. Pandey, H. Kontani, D. S. Hirashima, R. Arita, and H. Aoki. Spin Hall effect in iron-based superconductors: A Dirac-point effect. Phys. Rev. B \textbf{86}, 060507(R) (2012).

\bibitem{kontani_prl_09} H. Kontani, J. Goryo, and D. S. Hirashima. Intrinsic Spin Hall Effect in the s-Wave Superconducting State: Analysis of the Rashba Model. Phys. Rev. Lett. \textbf{102}, 086602 (2009).

\bibitem{sengupta_prb_06} K. Sengupta, R. Roy, and M. Maiti. Spin Hall effect in triplet chiral superconductors and graphene. Phys. Rev. B \textbf{74}, 094505 (2006).

\bibitem{bergeret_prb_16} F. S. Bergeret and I. V. Tokatly. Manifestation of extrinsic spin Hall effect in superconducting structures: Nondissipative magnetoelectric effects. Phys. Rev. B \textbf{94}, 180502(R) (2016).

\bibitem{malshukov_prb_10} A. G. Mal’shukov, S. Sadjina, and A. Brataas. Inverse spin Hall effect in superconductor/normal-metal/superconductor Josephson junctions. Phys. Rev. B \textbf{81}, 060502(R) (2010).

\bibitem{malshukov_prb_17} A. G. Mal'shukov. Supercurrent generation by spin injection in an 
s-wave superconductor–Rashba metal bilayer. Phys. Rev. B \textbf{95}, 064517 (2017).

\bibitem{wakamura_natmat_15} T. Wakamura, H. Akaike,	Y. Omori,	Y. Niimi,	S. Takahashi,	A. Fujimaki,	S. Maekawa, and Y. Otani. Quasiparticle-mediated spin Hall effect in a superconductor. Nat. Mater. \textbf{14}, 675 (2015)

\bibitem{zhihong_cpb_12} Y. Zhi-Hong \etal, Chin. Phys. B \textbf{21}, 057402 (2012).

\bibitem{malshukovchu} A. G. Mal'Shukov and C. S. Chu, Phys. Rev. B \textbf{78}, 104503 (2008); \textit{ibid} \textbf{84}, 054520 (2011). 





\end{thebibliography}
\end{document}